# Agency, Affordances, and Enculturation of Augmentation Technologies

By Ann Hill Duin and Isabel Pedersen

## Overview

Augmentation technologies are undergoing a process of enculturation due to many factors, one being the rise of artificial intelligence (AI), or what the World Intellectual Property Organization (WIPO) terms the “AI wave” or “AI boom.” Chapter 3 focuses critical attention on the hyped assumption that sophisticated, emergent, and embodied augmentation technologies will improve lives, literacy, cultures, arts, economies, and social contexts. The chapter begins by discussing the problem of ambiguity with AI terminology, which it aids with a description of the WIPO Categorization of AI Technologies Scheme. It then draws on media and communication studies to explore concepts such as agents, agency, power, and agentive relationships between humans and robots. The chapter focuses on the development of non-human agents in industry as a critical factor in the rise of augmentation technologies. It looks at how marketing communication enculturates future users to adopt and adapt to the technology. Scholars are charting the significant ways that people are drawn further into commercial digital landscapes, such as the Metaverse concept, in post-internet society. It concludes by examining recent claims concerning the Metaverse and augmented reality.

## Key Questions

1. How is AI emerging to generate augmentation technologies?
2. Amid emerging augmentation technologies, how are humans and AI technologies framed as agents with agency?
3. How is the massive corporate backing of Metaverse as the next phase of the internet changing the design, adoption and adaptation of augmentation technologies?

4. How are inventors and stakeholders envisioning cognitive, sensory, emotional, and physical enhancement?

## Chapter 3 Links

Throughout the chapter, we refer to articles, videos, and reports which can be found in a related chapter collection at Fabric of Digital Life called *Agency, affordances, and enculturation of augmentation technologies* (Duin & Pedersen, 2022). You can find a link to this collection at https://fabricofdigitallife.com.

## Introduction

Susan B. Anthony declared in 1896 that bicycling had “done more for the emancipation of women than anything else in the world” (Bly, *New York World*, p. 10). Sarah Hallenbeck (2012) begins her article on user agency in technical communication with this quote, noting how Anthony’s statement was “characteristic of the cultural moment in which she was writing—in which a bicycle craze had infected people in nearly every class, and the demand for ‘the steed which never tires’ was so high . . . that many bicycle factories operated around the clock,” with women celebrating the transformative power of this new technology that increased “mobility, independence, and physical health and strength” (p. 290). However, “moral conservatives warned that bicycling, as well as the changes in attire that it required, would ‘unsex’ women riders” (p. 291). Despite these warnings, women’s riding only increased, and bicycle manufacturers began to add features in support of “the embodied practices of women themselves,” constituting a “regendering of the bicycle, whereby the once masculine object now developed strong associations with female riders” (p. 291). Hallenbeck investigates how these early women riders, “as users excluded from official avenues of design and manufacture” (p. 291), laid claim to this technology not originally designed or intended for their use. She complicates the notion that technology “becomes a stable and static object once it enters into common use" (p. 291). To do so, she extends her focus to contexts of use, highlighting what sociologist Andrew Pickering (1995) called a "dance of agency," noting that in this configuration, "users are not merely passive recipients of producers' expert knowledge and inventive genius but also shapers of the cultural and symbolic meanings attached to the technologies with which they interact" (Hallenbeck, p. 291, citing Oudshoorn & Pinch, 2003).

More recently, Rosi Braidotti (2021) contends that "the posthuman turn can result in a renewal of subjectivities and practices by situating feminist analyses productively in the present" (p. 106). Over the past decade, TPC scholars have embraced the social justice turn by "recognizing the injustices and oppressive systems embedded in our work as technical communicators" (p. 133). Rebecca Walton states in an interview that "intersectionality is an important part of our thinking right now because it attunes us to the compounding nature of oppressions for people who are multiply marginalized" (Walton, Moore, & Jones, 2019, p. xxiii). An ethical framework for augmentation technologies needs to revise older patterns of thinking and acting in the field. Walton et al. (2019) contend that "Social injustices require coalitional action, collective thinking, and a commitment to understanding difference that is not necessarily demanded by other technical communication problems" (p. 1).

Complicating our field is the fact that we are in the midst of an augmentation technology craze fueled partly by a blind trust in artificial intelligence. The WIPO names the phenomenon an "AI wave" (p. 19), calling it a "good time to take a close look at the state of research and exploitation of AI technologies" (p. 20). Bernard Marr (2020), writing for *Forbes*, identifies AI as "undoubtedly one of the biggest tech trends at the moment, and during 2021 it will become an even more valuable tool for helping us to interpret and understand the world around us. The volume of data we are collecting on healthcare, infection rates, and the success of measures we take to prevent the spread of infection will continue to increase. This means that machine learning algorithms will become better informed and increasingly sophisticated in the solutions they uncover for us." He continues by citing robotics as a means to reduce human labor costs and balance uncertainty and customer demand, and citing extended reality (XR), virtual reality (VR), and augmented reality (AR) as tools to help us avoid potentially dangerous situations, e.g., allowing medical examinations and diagnosis to be carried out remotely, providing real-time warnings about virus spread, and guiding us through the challenging world that we face. Likewise, Brian Chen (2021), writing for *The New York Times*, highlights our embrace of augmentation technologies to interact with others and with physical worlds now less safe given the pandemic. And XPRIZE stories showcase humans breaking boundaries with myriad uses of augmentation technology; for example, novelist Kazuo Ishiguro says, "When you get to the point where you can say that [a] person is actually intellectually or physically superior to another person because you have removed certain possibilities for that person getting ill . . . or because they're enhanced in other ways, that has enormous implications for very basic values that we have" (XPRIZE, 2021).

## AI Technologies and Enculturation

In previous chapters, we have explained that the rise of AI, in and of itself, generates the rise of augmentation technologies. We argue that augmentation technologies are undergoing a process of normalization due to massive transformations in AI leading to the enculturation of both. However, definitions for AI are ambiguous. One of the issues TPC professionals are facing is the lack of clarity about what artificial intelligence (AI) means:

> AI is "the ability of certain machines to do things that people are inclined to call intelligent."
>
> Interview with Marvin Minsky in *The New Yorker* (Bernstein, 1981)

> AI can be defined as “Systems that think like humans, systems that act like humans, systems that think rationally, systems that act rationally.”
> Stuart Rusell and Peter Norvig in *Artificial Intelligence: A Modern Approach* (Russell & Norvig, 1995)

> AI is “the 60-year-old quest to make machines capable of mental or physical tasks seen as emblematic of human or animal intelligence.”
> Tim Simonite, *Wired* Magazine (Simonite, 2019).

> Artificial general intelligence (AGI) can be defined as “highly autonomous systems that outperform humans at most economically valuable work.”
> OpenAI Charter (OpenAI*,* 2018)

To deal with shifting concepts in this chapter, we categorize AI technology under three terms for the sake of discussion in this book. The terminology is drawn from the Association for Computing Machinery (ACM) Computing Classification Scheme, as quoted by the World Intellectual Property Organization (2019). WIPO explains the intent for better clarity: “this scheme has the advantage of providing a clear analytical framework for the report and the presentation of the evolution of AI technologies over time” (p. 25).

Figure 3.1 illustrates the category called *AI techniques*, which are “advanced forms of statistical and mathematical models, such as machine learning, fuzzy logic and expert systems, allowing the computation of tasks typically performed by humans” (p. 25). AI techniques should be seen as the technical means to create *AI functions* within applications.

*Figure 3.1* AI Techniques (WIPO, 2019, p. 24). Creative Commons Attribution 3.0 IGO license. https://creativecommons.org/licenses/by/3.0/igo/

Figure 3.2 illustrates *AI functional applications*, which include “functions such as speech [processing, predictive analytics] or computer vision which can be realized using one or more AI techniques” (p. 25). Computer vision informs and enhances *augmented reality*, which is already a developed computing field. But this infusion of AI techniques makes augmented reality much more sophisticated than before.

*Figure 3.2* AI functional applications (WIPO 2019, p. 26). Creative Commons Attribution 3.0 IGO license. https://creativecommons.org/licenses/by/3.0/igo/

Figure 3.3 illustrates *AI application fields*, which include the many “fields, areas or disciplines where AI techniques or functional applications may find application, such as transportation, agriculture or life and medical sciences” (p. 25).

F*igure 3.3* AI application fields (WIPO 2019, p. 27). Creative Commons Attribution 3.0 IGO license. https://creativecommons.org/licenses/by/3.0/igo/

Words describing AI often appear as common buzzwords in popular technology discourses. But one can better imagine AI technologies in the stratified WIPO scheme whereby concepts such as *smart cities* are constituted through a hierarchy of the categories (see Figure 3.4). For example, *AI techniques* can be combined to create *AI functional applications,* which lead to fields like *Networks*. Networks help to constitute Internet of Things (IoT), smart cities, and social networks in working structures where humans interact with interfaces. The figure below illustrates the scheme.

*Figure 3.4* Explanation of WIPO Categorization of AI Technologies Scheme.

## Agency in the Digital Age amid Human and Machine Cultures

This book starts with a commitment to reconceptualize augmentation technologies through a human-centric, ethical framework, acknowledging the rich assemblages within which interactions occur between humans and non-human agents. In the first chapter, we point to the hyped rhetorics of betterment that often accompany these enhancement technologies, which sometimes obscure our ability to assess their use. Taking it a step further, Zizi Papacharissi (2019) writes of the theme of augmentics, “if we imagine and then design technology as something that augments or enhances our human powers, then we are bound to wind up with bizarre superhero complexes, to say the least, and grossly underestimate *techne*” (p. 7). She does, however, move toward a more ameliorative vision for augmentics. She writes, “we have the opportunity to become different, and with this opportunity comes another possibility, perhaps an obligation: to reimagine how we do things; to do things differently; to not fall into the trap of reproducing what we already have” (p. 7).

In communication and media studies as well as technical communication theory, agency is associated with people, with users. To use a bicycle, an agent (the rider) must have the agency or capacity to do so. According to Kris Cohen (2022), “expressed simply, an agent is one who acts. The power granted or effected through that action is the quality of agency. . . . Any definition of agency is complicated from the start by disciplinary differences in its conceptualization and use.” AI, sociology, and history disciplines cite “social agents” as empirical or quasi-empirical entities; political and feminist theories view agency as a tool aligned with subjectivity and marginalization; and philosophical work on autonomy sees agency as a relation of self to self. According to Michel Foucault (1982), agency is a concept based on power relations; for Martin Heidegger (1977), agency is defined within new technologies of representation, e.g., within the “collective” of the agent (set of scientists, observers); and for Marshall McLuhan (1964), technology acts on human agents to extend perception and cognition, with the clear danger of being blinded to the ways technologies impact our existence. Karlyn Kohrs Campbell (2005) defines agency as “polysemic and ambiguous, a term that can refer to invention, strategies, authorship, institutional power, identity, subjectivity, practices, and subject positions” (p. 1). Cohen (2022) states that all of this literature denotes the effective agent as “the powerful self, the assertive self, the self who is able to impose her will, if only over the terms and conditions of her own life.”

Bruno Latour’s (1999) Actor-Network Theory (ANT) denotes agency as a matter of attribution; i.e., that things have agency as a result of their position within a network, and that everything exists in constantly shifting networks of relationships. Molly Kessler and Scott Graham (2018) articulate this position within technical communication: “ANT has been primarily used in efforts to understand (1) the coarticulation of humans and nonhumans institutions working together to accomplish a goal or set of goals, and (2) the complexity and density of these networks, and (3) the nature of power and agency as network properties (Graham & Herndl, 2011; Miettinen, 1999; Potts & Jones, 2011; Spinuzzi, 2007; Winsor, 2006)” (p. 123). ANT scholarship through these recent decades has deepened our understanding of the social lives of technologies, elucidating how agency is embedded in a broader sociocultural landscape (Neff & Nagy, 2019), with specific studies such as Kessler and Graham’s (2018) examination of the embedding of agency in prescription drug labels.

Advancements in recent technologies challenge these fundamental notions of agency, as “tools and techniques including machine learning, artificial intelligence, and chatbots may be capable of exercising complex ‘agentic’ behaviors” (Neff & Nagy, 2019, p. 97). As discussed in Chapter 1, robotic companions now read human emotions, and chatbots evoke reactions to help people make decisions or cope with anxiety. In Chapter 2, we classified this interactivity as emotional enhancement. Calling for a better definition of agency amid these “complex, technologically mediated interactions,” Neff and Nagy ask, “Does the agency of the people and, increasingly, of things that we choose to communicate with matter?” (p. 97). They employ Albert Bandura’s (2006) social-cognitive psychological theory of human agency to “reframe and redefine agency in terms of how and when people interact with complex technological systems. . . . Being an agent means that people can exert intentional influence over their mental processes and physical actions to achieve their desired outcomes” (p. 98). In short, people delegate tasks to augmentation technologies as their personal assistants and companions, resulting in a symbiotic relationship. Using Bandura’s theory, Neff and Nagy link agency to intentionality, forethought, and self-reflectiveness. Collective agency indicates shared beliefs and goals, and proxy agency involves other people or tools that can help one achieve goals. They propose the term “symbiotic agency” as a specific form of proxy agency, i.e., when people interact with technologies:

> They exercise proxy agency through a technologically mediated mode referring to the entanglement of human and non-human agencies. For symbiosis, similar to entanglement, implies an obligated relationship—symbionts are completely dependent on each other for survival, it can be considered a proxy agentic relationship that may provide different benefits for the recipients. (p. 102).

People can benefit from, not be affected by, or be harmed by symbiotic relationships. They note how the “Uncanny Valley Effect” (Reichardt, 1978) “documents the discomfort that people feel when robotic agents are seen as having a too humanlike appearance. . . . When ascribing agency to complex technologies, such as chatbots, robots, or other AI applications, people project their emotions to these entities and try to find explanations for their ‘behaviors’ (Darling, 2014)” (p.103). Neff and Nagy explain that these relationships can also be parasitic rather than symbiotic, noting “users may also feel that they are increasingly getting ‘addicted to’ or ‘controlled by’ technologies due to the perceived parasitic attributes of these tools.” (p. 103).

We contend that augmentation technologies elicit symbiotic agency; any lines between what users do and what technologies do are now forever blurred. As augmentation technologies listen in on conversations and read thoughts, they learn and shape future meaning and responses, resulting in increasingly intimate relationships. “By redefining human–technology interaction as a unique form of proxy agency, [we] can investigate both the human agentic properties and the technological attributes of what it means to be a networked self” (Neff & Nagy, 2019, p. 104).

In short, computers adapt to humans. The affordances for directing computers include speech, gestures, eye gaze, and even human electrophysiological signals (Raisamo et al., 2019). We gain direction while driving that includes auditory as well as haptic smartwatch notification to make sure to turn at the appropriate time. Roope Raisamo et al. discuss such information flow and associated technologies in wearable augmentation: “Numerous sensors and cloud data provide information, artificial intelligence filters it and it is presented in easy-to-understand ways to support human cognition in a timely manner. Physical tools or robots enable action in and changes to the environment” (p. 136). Moving forward, technological ability to sense humans will intensify as the Internet of the Body evolves. Technologists are orienting development toward human-centric values: “only a system that is able to extend its capabilities to perceive and adapt to human actions, intentions, and emotional states, can be considered a truly smart system” (Fernandes et al., 2022).

In effect, we have created human and machine cultures, the posthuman complete with cognitive assemblage, technical agency, and human interactions. N. Katherine Hayles (2002), prominent posthuman theorist, states that “Agency still exists, but for the posthuman it becomes a distributed function” (p. 319). She clarifies our lived experience as follows: “Living in a technologically engineered and information-rich environment brings with it associated shifts in habits, postures, enactments, perceptions—in short, changes in the experiences that constitute the dynamic lifeworld we inhabit as embodied creatures” (p. 299). Reaffirming her position in a later book (2017), she explains post-digital human and non-human collaboration in terms of “complex human technical assemblages in which cognition and decision making powers are distributed throughout the system. I call the latter cognitive assemblages” (p. 4).

## Non-human Agents Become Enculturated in Industry Sectors

How are non-human agents emerging in industry sectors in terms of collective agency? How is ethical alignment or shared beliefs and goals appearing in the discourse of emergence, if at all? In a recent exploratory study to identify the nature of emergent relationships between human and non-human agents, we analyzed how working assemblages of activity are taking shape. The concentration was on AI technologies marketed for learners in networked learning environments because we wanted to better understand how AI is proposed to augment humans in this context (Pedersen & Duin, 2022). We looked at corporate advertisements for new educational technologies, as well as ones that simply claimed to have an educational function. Several were still in corporate pre-release research and development hyping future products to help raise funds and awareness for startups. We did not include university or government research because the goal for this study was to target products that will enter classrooms and people’s homes, ones that are closer to release. In Chapter 2, we discussed claims about augmentation that drive adoption and adaptation. Our intent with this study was informed by the same research goal, to reveal both affordances and consequences of these relationships. We used Fabric of Digital Life (https://fabricofdigitallife.com/) metadata to identify a dataset of 17 video advertisements for AI inventions. It covered wearables, mobile, and robotic devices. Marketing materials also included sales copy about AI and how an AI agent, for example, could improve human skills. We applied a critical discourse analysis framework to reveal these persuasive statements in the advertising. We analyzed the way that children were visually depicted with robots and human teachers in order to understand how human agents and AI agents are promoted in collaborative cognitive assemblages. At the same time, robot actions and behaviors (signifiers) are realized through composites of sensors, trackers, cameras, moving mechanical parts, and AI technologies, etc., built to operate as if a unified agent. This information can be tracked through the technology keyword category. Fabric provides grounds for analysis through the metadata. Styles of communication change with each new technical advancement signaling the continuous possibility of novel augmentation.

The advertisements also frame agentive relationships through values expressed to promote their adoption. A good example is Little Sophia by Hanson Robotics, a spin-off of the famous Sophia robot (see Figure 3.5). Little Sophia is a consumer-grade device still in pre-release with the potential to be used in homes and classrooms. The suggested scenarios for Little Sophia appear in the ad copy: “with Little Sophia's software, and included tutorials through Hanson’s AI Academy, she is a unique programmable, educational companion for kids, inspiring children to learn through a safe, interactive, human-robot experience” (Hanson Robotics, 2019). Little Sophia is also always enveloped in Sophia’s fame as a much more developed AI agent/celebrity that takes every opportunity to promote her value. It is unlikely that any household will have a Sophia robot for decades. It is much more likely that students, workers, and consumers will be involved in adapting to functionality for robots on the scale of Little Sophia that will eventually progress to that of Sophia.

Overall, the study found imbalances in proposed human-nonhuman relationships, “that [the] advertising promotes corporate products while also promoting idealized social practices for human-computer interaction and human-robot interaction in learning contexts. Using AI to automate relationships between students and teachers frames AI systems as authorities in both robot and non-robot platforms, blurring and minimizing student and instructor agency in learning environments” (Pedersen & Duin, 2022, p. 1).

*Figure 3.5* Sophia at the AI for Good Global Summit 2018. Photography by ITU Pictures from Geneva, Switzerland. Creative Commons Attribution 2.0 Generic license. https://creativecommons.org/licenses/by/2.0/

We prepared Table 3.1 below, Rhetorical Propositions about AI and Augmentation Technology in Marketing Videos, to exemplify some of the marketing claims by corporate rhetors. Column 1 identifies the type of augmentation hardware platform about which claims are made. Column 2, “Video artifacts and main augmentation,” lists each video and the central augmentation it promises to provide, many of which are abstract value propositions. Column 3, “Persuasive claims about augmentation made by companies,” includes quotes from the videos or textual ad copy surrounding the artifacts. Column 4 lists technological keywords used to substantiate those claims.

Each row provides a snapshot of idealized scenarios for an augmentation aligned with the technologies proposed to fulfill them. Each artifact serves to enculturate consumers to adopt a product and adapt to it. For example, Waverly Labs (2019) promises to be working toward “A world without language barriers,” a significant augmentation, and we report in column 4 exactly which technologies are being offered to fulfill such a claim. Composites of embodied devices (wearables: earbuds, microphones), AI technologies (neural networks, speech recognition), and internet infrastructure (cloud computing, wireless connectivity) are implicated in fulfilling these rhetorical claims.

TPC practitioners need greater awareness of these components as they rapidly emerge to support their functional use in homes, schools, and/or work scenarios. Robots operating in ambient spaces around humans form cognitive assemblages and increased emotional enhancements. For example, many of the robots use pre-built facial expressions presumably to establish empathetic communication channels by mimicking human body language. TPC will be involved in writing these scripts for robots to legitimize the messages and constitute robots as agents. However, awareness of the affordances of these developing technologies is and will be key (for additional discussion, see Ayanoglu & Duarte, 2019).

| Hardware Platform | Video artifacts and main augmentation | Persuasive claims about augmentation made by companies | Technological keywords |
|---|---|---|---|
| wearables | | | |
| | Ambassador by Waverly Labs<br>Company: Waverly Labs, 2019<br><br>augmentation:<br>Learn new languages | “imagine being able to snap your fingers and become fluent in 20 languages”<br><br>“A world without language barriers” | Artificial Intelligence (AI), Neural Networks, Speech Recognition, Translators, Wearable Translators, Smart Earbuds, Microphones, Speakers, Real-Time, Cloud Computing, Smartphone Applications, Wireless Connectivity |
| | FOVE VR Headset: Tracks Subtle Eye Movements in Virtual Reality<br>Company: Fove, Inc., 2015<br><br>augmentation:<br>Interact in virtual worlds | “We are on a mission to unlock the essence of reality in virtual worlds”<br><br>“Enables new forms of expressions, communication and movement” | Virtual Reality (VR), Virtual Worlds, Assistive Technologies, Artificial Intelligence (AI), Headsets, Head Mounted Displays (HMD), Cameras, Eye Tracking, Infrared Tracking Systems, High Resolution |
| | | | Display, Natural User Interface (NUI) |
| robots | | | |
| | Little Sophia by Hanson Robotics<br>Company: Hanson Robotics, 2019<br><br>augmentation:<br>Inspire imagination and learning | “She can take you on wild adventures” | Robotics, Humanoid Robots, Social Robots, Human-Robot Interaction (HRI), Artificial Intelligence (AI), Facial Recognition, Pre-Built Facial Expressions, Cameras, Microphones, Speakers, Smartphone Applications, Open Source, Wireless Connectivity |
| | Xiaoyou robots made by Canbot<br>Company: Canbot, 2018<br><br>augmentation:<br>Automate teaching, learning, and living | “Robot for your better life” | Robotics, Humanoid Robots, Social Robots, Human-Robot Interaction (HRI), Artificial Intelligence (AI), Natural Language Processing (NLP), Voice Recognition, Intelligent Home Control Systems, Smart Homes, Internet of Things (IoT), Tablets, Cameras, Microphones, Speakers, Video Calling, Sensors, Obstacle Detection, Remote Monitoring, Wireless Connectivity |
| mobile | | | |
| | Meet AI XPRIZE Semifinalist Iris.AI<br>Company: Iris.AI, 2020<br><br>Augmentation:<br>Keep up with research, become efficient, overcome information overload | “Iris is an AI science assistant to help researchers do their work” | Artificial Intelligence (AI), Algorithms, Machine Learning, AI Research Assistants, Text Generators, Text Summarizer, Data Collection, Data Extraction |
| | MondlyAR - learn languages in augmented reality | “The app creates a bridge between two worlds: your environment and a virtual universe” | Augmented Reality (AR), Mobile Augmented Reality, Artificial Intelligence (AI), Speech |
| | Company: Mondly Languages, 2018<br><br>Augmentation:<br>Learn new languages | "your personal language learning assistant" | Recognition, Virtual Assistants, Virtual Worlds, Cameras, Microphones, Speakers, Smartphone Applications |

*Table 3.1* Rhetorical Propositions about AI and Augmentation Technology in Marketing Videos.

Stakeholders project agency *upon* non-human agents. Consider the Amelia Integrated Platform (2022), through which conversational AI/Amelia, a non-human agent, takes on the role of "digital employee" to deliver "the best elements of human interaction—conversation, expression, emotion and understanding—to user experiences every day, driving deeper connections and greater business value." Amelia's agency, both rhetorically and ethically provocative/hyped/problematic, inflates her role as an imagined employee. A recent article describing 50 robots that now work as cleaners, but also as surgeons and physiotherapists, at Singapore's high-tech hospital also refers to these as "AI-powered employees" (Cairns & Tham, 2021). As technical and professional communicators, our task is to understand this evolution, positioning ourselves in collaboration (human-autonomy teaming), with focus on human-centered design. We further discuss digital employees in Chapter 7.

## Metaverse, Augmentation Technology, and Virtual Worlds

Corporate playbooks contain initiatives to drive higher revenues by changing the way people use their devices. Many writers are pointing out how people are drawn further into commercial digital landscapes in post-internet society (Crawford, 2021; Mosco, 2017; Andrejevic, 2020). In Chapter 2, we discuss justifications for augmentation, including the goal to enhance the senses to construct an enhanced reality. One dramatic pitch by several "big tech" companies is the Metaverse trope. Definitions for the term "Metaverse" have not solidified. One article names it "a parallel digital reality in which users play and work—and can buy and sell in cryptocurrencies" ("Big Tech's Supersized Ambitions," 2022). Another calls it an "emerging virtual market which could, depending on whom you ask, ultimately generate revenues of between $1trn and $30trn" (MacManus, 2022). More often it is pondered about: "Will it be an all-consuming futuristic world of virtual reality, avatars, oceanside mansions and other online razzmatazz that will make the real world a dull place by comparison? Or will it simply be a richer, more immersive version of what already exists today?" ("Schumpeter: Lords of the Metaverse," 2021).

In October 2021, the Metaverse became wildly popular when "Mark Zuckerberg renamed Facebook *Meta* and described humankind's new future in virtual worlds" ("Big Tech's Supersized Ambitions," 2022). In some ways, the Metaverse is an incredibly hyped research project. One article reports that "somewhere between 5% and 20% of the tech giant's massive R&D spending goes towards what, for the purposes of this article, we are calling 'frontier technologies': the metaverse, autonomous vehicles, health care, space, robotics, fintech, crypto and quantum computing" ("Big Tech's Private Passions," 2022). Yet, the Metaverse does not exist in any real capacity; it is a dream: "The metaverse is for the foreseeable future quite literally science fiction: a fully immersive, persistent virtual world where, with the help of high-tech goggles and other kit, people interact, work and play via online avatars of their real-world selves" ("Building a Metaverse with Chinese Characteristics," 2022). Significant sensory enhancements would need to be adopted in order to achieve such a virtual world. Augmentation technologies would provide the means to traverse these interconnected virtual worlds, amounting to the next-generation 3D internet.

As one example, watch Joanna Stern, Senior Personal Technology Columnist for *The Wall Street Journal*, in her video *Trapped in the Metaverse: Here's What 24 Hours in VR Feels Like*. She confronts the fantastical claims made about the medium with realistic questions such as "Will we sleep in VR?" "Will we have multiple avatars?" and "How will we navigate between working and socializing in metaverse virtual worlds?" The video gives TPCs a snapshot of usability affordances as well as drawbacks when people are asked to use a VR headset for whole days. It illustrates how multiple human–computer interaction styles are evolving to help people communicate with each other and their own digital information.

In response to the hype, augmentation technology companies have entered the fray, using the Metaverse as a persuasive concept to excite, enculturate, and innovate. Metaverse terminology helps simplify complex technology descriptions, and works to sell people on the idea of enhancement as an alluring and realizable future.

Moreover, the Metaverse concept taps longstanding Big Tech corporate ambitions. Tracing a few of them explains its trajectory and how it will contextualize augmentation technology in future through the themes that are dominating this call for adoption. On July 20th, 2009, *New York Times* writer Richard MacManus made one of the many early predictions of a Metaverse in his article, "The Wearable Internet Will Blow Mobile Phones Away." He explained why people should eschew clunky mobile phones and adopt wearable components:

> The Web as we know it today is full of manual steps, such as visiting websites and searching for information. In 10 years time we'd hope that the Web of Data would be much better realized. . . . You can see the power of this as a next generation Internet interface, as it removes several manual steps from the process of receiving relevant, contextual information about something or someone. . . . There are also sensors in the objects—for example the book has a barcode that, in combination with the wearable device, will pull down data from Amazon.com via the Web. . . . [We] are very excited about next-generation Internet interfaces, such as augmented

> reality and so-called cross reality. These wearable devices strike me as being the most impressive future Web interface that I've seen in a while.

MacManus’s 2009 call to design this future internet hangs on interoperability, the desire for more efficiency when interacting with devices and fewer “manual steps” (MacManus, 2009). Now more than a decade later, a “Web of Data” has emerged, and users are very much embedded in algorithmic cultures “receiving relevant, contextual information about something or someone.” MacManus also predicted an Internet of Things with the idea of “sensors in the objects” leading to more integration with corporate spheres, such as Amazon. Interestingly, Amazon was worth $59.72 billion in market cap in 2009; now it is worth $1,598 trillion in 2022, justifying its vision (Market Capitalization of Amazon, 2022). Amazon wildly exceeded expectations as an internet and AI company. In 2009, MacManus also plugged immersive media (sensory enhancement) such as “augmented reality” to provide the means to interact more efficiently with his envisioned internet. He argued for more efficiency, automation, and embodiment for internet interactivity, which aligns with our arguments for this book concerning augmentation technology emergence and its promotion.

Fast forward to 2022, and Richard MacManus now writes about the Metaverse’s potential for creative content collaboration (MacManus, 2022). He describes “Meta’s vision of an interoperable set of competing virtual worlds,” but he also describes Nvidia’s “Omniverse [which] has officially turned into a content creation platform for virtual worlds on the consumer internet. Nvidia is welcoming ‘millions of individual creators and artists’ onto its platform, to build for the next generation internet” (MacManus, 2022). Rather than HTML, content creators will write Universal Scene Description (USD); all of Omniverse is built on top of Pixar’s open source Universal Scene Description (Pixar, 2022), which Nvidia has pitched as the “HTML of 3D worlds” (MacManus, 2022).

So, how will metaverses be built? Technical and professional communicators can read about USD and shared virtual worlds. Michael Kass, Senior Distinguished Engineer at Nvidia Omniverse, describes USD and writes, “at the core of any true 3D web or the open Metaverse, we need an open, powerful, flexible and efficient way to describe the shared virtual world” (Kass, 2021). *Computer Graphics World* explains how Pixar developed USD over the past 25 years, stating “we have needed ways to describe the 3D scenes we are synthesizing in a way that is mathematical enough for computers to understand, but which is also understandable and manipulatable by technologists *and* artists” (Grassia & ElKoura, 2020).

## Metaverse and Augmented Reality

Inventors and stakeholders are envisioning cognitive, sensory, emotional, and physical enhancement for the same reasons they did a decade ago, albeit with more intensity than ever before. The 2022 CES show, already a venue for hyped announcements, featured several augmentation technologies under the Metaverse banner. InWith Corporation claims to provide “potentially the most advanced platform for viewing the coming Metaverse.” Rather than announce yet another phone-based application, InWith Corporation explains that the platform would be hosted on the user’s eyeballs: “an electronic soft contact lens platform designed for the masses to wear comfortably, enabling an easy transition from real-world to Metaverse, at will” (InWith at CES 2022, 2022). In one broadcast by CNET, the concept is broken down in simple terms promoting the idea that users will be able to flip from real-world interaction (augmented reality AR) to virtual immersion (virtual reality VR) much more seamlessly by wearing contact lenses, leading to significant sensory enhancement (Altman, 2022).

We have defined augmentation technologies as those that enhance human capabilities or productivity by adding to the body in the name of efficiency and automation. InWith Corporation soft contacts and platform are offered as an efficient means to access the Metaverse (a concept that does not yet exist in structural terms as we have discussed). Most portraits of Metaverse experiences promote the duality of AR and VR, that users will glide into and out of virtual worlds with ease. However, actually achieving that experience usually means adopting much clunkier hardware options with smart glasses for AR and larger headsets for VR, like Oculus Quest 2. The InWith soft contacts serve as an exemplar for evolving user expectations because they would provide an easier, more seamless, means to access information.

Fabric has been used to chart the smart contact lens industry (also known as bionic contact lenses) with artifacts dating as far back as 2010. One artifact from 2012 (see Figure 3.6) features the famous futurist Michio Kaku, who says that the “internet will be in our contact lens” and that we will “simply blink and be online,” and that this will assist with communication, identification, and language translation. A decade of research and technical innovation has contributed to this emergent augmentation technology. Studying smart contact lens artifacts along a timeline in Fabric helps to reveal how the InWith contact lens is the first to be used in combination with Metaverse in 2022. The convergence of the smart contact lens industry and Metaverse suggests a generative enculturation process at work. It signals how companies are using Metaverse hype to encourage adoption and eventually adaptation of both concepts. It also provides technological grounding for the infrastructure that will be needed to make smart contact lenses usable.

*Figure 3.6* Smart Contact Lens Timeline featuring a 2012 video artifact. Used with permission of Isabel Pedersen.

Another major hurdle to actually creating a Metaverse is current internet infrastructure. Both authors have been researching AR as an important medium of communication for the TPC industry along with many collaborators (Duin et al., 2020; Pedersen et al., 2021). AR is a method of human–computer interaction that blends virtual representations with physical spaces, seeking to augment, mix, or shift reality (Azuma, 1997; Pedersen, 2013). After undergoing 20 years of emergence, AR is predicted to progress to become one part of a full-scale ambient platform, helping to constitute what is recognized as a Metaverse. The transformative quality of AR is that it combines virtual components with human sensory modalities (e.g., sight, sound, touch), meaning that it can enhance the first-person point of view of the user with the potential to achieve a significant sense of virtual presence.

Metaverse technologies will have to achieve interoperability for data exchange over spatial computing platforms. One technology that has been evolving for some time is AR cloud, a speculative plan to create a shared multi-user environment, a global augmented reality, and "a real time spatial map of the world" (AWE, 2018). For example, the AR cloud requires a persistent 3D digital copy of the world, which means that if a developer creates a virtual overlay for a real space, it will remain accessible to anyone across different devices and platforms. As an imagined technology future, an AR cloud could both effect positive societal change (e.g., Open AR Cloud Association) and result in greater risk through further conditions of surveillance, corporate control, or harms to humans.

Immersive technologies are already being commandeered to enable social and cultural change for the better. Lili Yan, Mckay Colleni, and Breanne Litts (2020) argue that “immersive technology has the potential to represent the concept of the Anthropocene by engaging the user with the messages rhetorically presented.” They further explain the utility of AR:

> The concept of the Anthropocene criticizes the anthropocentric human-nonhuman relationship by focusing on the interconnectedness between human and nonhuman. In

> this sense, the Anthropocene is a deeply cultural phenomenon . . . a small number of AR applications have emerged to support the understanding of the Anthropocene, the key tenets of which are usually rhetorically communicated. (p 109)

In short, reconceptualizing augmentation technologies through an ethical framework requires us to acknowledge and explore the ways that we are enculturated to adopt them. By doing so, we cultivate digital literacy (see Chapter 4); moreover, through strategic pedagogical use of Fabric of Digital Life artifacts, we increase understanding of AI techniques, functional applications, and AI application fields (see Chapter 6).